# AN ANT-BASED ALGORITHM WITH LOCAL OPTIMIZATION FOR COMMUNITY DETECTION IN LARGE-SCALE NETWORKS

DONGXIAO HE, JIE LIU, BO YANG, YUXIAO HUANG, DAYOU LIU[*], DI JIN

*College of Computer Science and Technology, Jilin University, Changchun, 130012, China*
*Key Laboratory of Symbolic Computation and Knowledge Engineering of Ministry of Education, Jilin University, Changchun 130012, China*
hedongxiaojlu@gmail.com
liu_jie@jlu.edu.cn
ybo@jlu.edu.cn
jameshuang1981@gmail.com
liudy@jlu.edu.cn
jindi.jlu@gmail.com

In this paper, we propose a multi-layer ant-based algorithm MABA, which detects communities from networks by means of locally optimizing modularity using individual ants. The basic version of MABA, namely SABA, combines a self-avoiding label propagation technique with a simulated annealing strategy for ant diffusion in networks. Once the communities are found by SABA, this method can be reapplied to a higher level network where each obtained community is regarded as a new vertex. The aforementioned process is repeated iteratively, and this corresponds to MABA. Thanks to the intrinsic multi-level nature of our algorithm, it possesses the potential ability to unfold multi-scale hierarchical structures. Furthermore, MABA has the ability that mitigates the resolution limit of modularity. The proposed MABA has been evaluated on both computer-generated benchmarks and widely used real-world networks, and has been compared with a set of competitive algorithms. Experimental results demonstrate that MABA is both effective and efficient (in near linear time with respect to the size of network) for discovering communities.

## 1. Introduction

Many complex systems in the real world exist in the form of networks, such as social networks, biological networks, Web networks, etc, which are collectively referred to as complex networks. The area of complex networks has attracted many researchers from different fields such as physics, mathematics, computer science, etc. While a considerable body of work addressed basic statistical properties of complex networks, such as the existence of "small world effect" [1] and the presence of "power laws" in the link distribution [2], another property has also been paid particular attention, that is, "community structure": the nodes in networks are often found to cluster into tightly-knit groups with a high density of within-group edges and a lower density of between-group edges [3].

The ability to detect community structure has a large amount of usefulness in many aspects [4]. For example, nodes belonging to the same community may have much more common features than those in different communities, which could be used to simplify the functional analysis of complex networks. Furthermore, community structure may provide insights in understanding some uncharacteristic properties of a complex network system. For instance, in the world wide web, community analysis has uncovered thematic clusters; in biochemical or neural networks, communities may be functional groups and separating the network into such groups could simplify functional analysis considerably.

In the past few years, the most popular method to detect communities in graphs consists in the optimization of a quality function, modularity $Q$ introduced by Newman and Girvan [5]. Modularity $Q$ gives a clear and precise definition of the characteristics of the acknowledged community and has had very successful application in practice. But it is still not free from shadows. In particular, the modularity maximization suffers from resolution limits [6], not being able to discern the quality of modules smaller than a certain size. Also, it exhibits extreme degeneracies [7] such that the globally maximum modularity partition is typically hidden among an exponential number of structurally dissimilar, high-modularity solutions. Nevertheless, it seems that modularity optimization is still an effective way to detect community structure in complex networks at the present time. There are some main reasons. Firstly, the so-called resolution limit problem of modularity can be mitigated by some deliberate solutions, such as designing a hierarchical algorithm with intrinsic multi-level nature [8]. Secondly, if the network is relatively small and contains only a few non-hierarchical and non-overlapping modular structures, the extreme degeneracy problem is less severe and modularity maximization methods are likely to perform well [7]. Last but not least, proper assignment of weights to the edges of a complex network can circumvent both the resolution limit and degeneracy problems associated with modularity [9].

---
[*] Corresponding author.

The search for the partition with maximal modularity is in general a great challenge since it was proved to be a nondeterministic polynomial time (NP)-complete problem [10]. Many heuristics relying on different approaches have been introduced to approximate the optimal partition, and some of them are able to find fairly good approximate solutions in a reasonable time. But there is still room for improvements in their performance, in terms of both effectiveness and efficiency.

For the above problem, a multi-layer ant-based algorithm, namely MABA, is proposed here. In the basic version of MABA, namely SABA, each ant locally optimizes modularity $Q$ through propagating the label of its current position (vertex) to some others guided by a simulated annealing strategy. The local optimization each ant does in each cycle is affected by the actions of the former ants, which can be regarded as the underlying interactive mechanism of the ant colony. With the proceeding of SABA, the local optimization actions of all the ants progressively aggregate, which makes the community structure of the network gradually presented. Now, the above process is the so-called SABA. For further improving the performance of our method, a thought of "layer and rule" is introduced to it, which extends SABA to MABA. To be specific, we firstly run SABA on the original network. Once the communities are obtained, SABA is reapplied to a higher level network, where each detected community is regarded as a new vertex and the sum of the weights of edges between any two communities as the weight between the new vertices. The aforementioned process is repeated iteratively until no further improvement on modularity can be achieved, which corresponds to MABA.

This new MABA has several important properties. Firstly, it can find community division of high quality with low computational complexity. Specifically, in terms of clustering quality, it is close or superior to the three excellent algorithms reported by the famous survey [11]. Meanwhile, in terms of computation time, it only takes a near linear time and thus is well suitable to deal with large networks. Secondly, MABA naturally incorporates a notion of hierarchy by using the strategy of "layer and rule", thus it can offer the hierarchical community structures with different resolutions. Last but not least, the so-called resolution limit of modularity [6] also seems to be mitigated due to the intrinsic multi-level nature of our algorithm.

## 2. Related Works

Over the last decade, lots of approaches have been proposed to analyze the community structures in complex networks, which adopt different types of principles and techniques in physics, mathematics, computer science, and so on. They mainly include: **divisive methods**, e.g. Girvan-Newman (GN) algorithm [3]; **modularity optimization methods**, e.g. Fast Newman (FN) algorithm [12], Simulated Annealing (SA) algorithm [13], Fast Unfolding Algorithm (FUA) [8]; **label passing methods**, e.g. Label Propagation Algorithm (LPA) [14], hub-based algorithms [15,16]; **dynamic methods**, e.g. Finding and Extracting Communities (FEC) [17], Infomap algorithm [18], Ronhovde and Nussinov (RN) algorithm [19]; and others. The interested readers can consult the excellent and comprehensive survey by Fortunato [20].

Moreover, there are also several ant-based algorithms (ABA) for community detection, which is especially focused on by us. Liu et al. considered the behaviors of each individual and described an ant colony clustering algorithm for automatically identifying social communities from email network [21]; and later, they improved their method in the calculation of distance between objects and the swarm similarity [22]. Sadi et al. [23,24] focused on ant colony optimization techniques to find cliques (or quasi-cliques) in the network, and assign these cliques (or quasi-cliques) as nodes in a reduced graph, and then use conventional community detection algorithms on this graph. Jin et al. [25] proposed a new ant colony optimization strategy based on Markov random walks theory. This method relies on the progressive strengthening of within-community links and the weakening of between-community links, which will make an underlying community structure of the network gradually become visible. However, to the best of our knowledge, all these above ant-based algorithms belong to heuristic method without explicit optimization objectives; meanwhile, they are computationally expensive and do not have the ability to deal with large-scale networks.

## 3. Algorithm

### 3.1. *Problem definition*

In order to assess the partition of a network into communities, an important quality metric, namely modularity $Q$, has been proposed by Newman and Girvan [5]. The idea of this modularity $Q$ is from the intuition that a network with community structure is different from a random network. Therefore, it can be defined as the difference between the fraction of edges that fall within communities and the expected value of the same quantity if edges fall at random without regard for the community structure.

Given an unweighted and undirected network $N = (V, E)$ and suppose the nodes are divided into communities such that node $i$ belongs to community $c_{r(i)}$ in which $r(i)$ denotes the label of node $i$, then modularity $Q$ is defined as

$$Q = \frac{1}{2m} \sum_{ij} \left( \left( A_{ij} - \frac{k_i k_j}{2m} \right) \times \delta(r(i), r(j)) \right). \tag{1}$$

Here $A = (A_{ij})_{n \times n}$ denotes the adjacency matrix of network $N$, in which $A_{ij} = 1$ if nodes $i$ and $j$ connect with each other, $A_{ij} = 0$ otherwise. The $\delta$ function $\delta(u,v)$ is equal to 1 if $u = v$ and 0 otherwise. The degree $k_i$ of any node $i$ is defined as $k_i = \sum_j A_{ij}$, and the total number of edges $m$ in this network is defined as $m = \frac{1}{2} \sum_{ij} A_{ij}$.

In order to make the local optimization of modularity $Q$ possible, we rewrite Eq. (1) into Eq. (2). Then, function $Q$ can be expressed as the sum of function $f$ of each vertex. Here, from the perspective of each vertex, function $f$ denotes the actual number of edges of a vertex within community, minus the expected value of the same quantity if edges fall at random without regard for the community structure. Thus, function $f$ can also be regarded as a quality metric for communities, which represents the same meaning as function $Q$ in terms of the local view of each vertex.

$$Q = \frac{1}{2m} \sum_i f(i), \quad f(i) = \sum_j \left( \left( A_{ij} - \frac{k_i k_j}{2m} \right) \times \delta(r(i), r(j)) \right). \tag{2}$$

Furthermore, from the analysis of [26], $\forall\, i \in V$, the global function $Q$ is monotone increasing with the local function $f$ of each vertex $i$. This also means that, if the variation of one vertex's label results in an increase of its own function $f$ (under the condition that the labels of all other vertices do not change), function $Q$ of the entire network will also increase at the same time. This property guarantees that optimizing the function $f$ from the angle of each vertex can finally fulfill the aim of optimizing modularity $Q$ for the entire network. In other words, the function $f$ can be taken as a good local measure of the quality of community division. As follows, we will propose a new method, the idea of which is to optimize the global function $Q$ by making each vertex optimize its local function $f$.

### 3.2. *The main idea*

Our algorithm is a multiple layer method based on a thought of "layer and rule". Its basic version is a single-layer ant-based algorithm (SABA) with a local optimization strategy, whose main idea is as follows. To begin with, it initializes each vertex as a community and randomly distributes some ants on the network. Thereafter, it will proceed in a number of cycles. In each cycle, each ant randomly crawls from one vertex to another, and tries to propagate the label of its current position to a certain neighbor.

Let the current position where this ant situates be vertex *i*, and the neighbor randomly selected by the ant be vertex *j* (which is treated as the next position). Then, this ant will decide whether vertex *j* should accept the label of vertex *i* as its new label according to a simulated annealing probability, purpose of which is to optimize the function *f*. In other words, this is indeed a type of simulated annealing strategy which allows vertex *j* to optimize the global function *Q* from its local perspective. This algorithm stops when there is no label changed by the ants, which can improve the modularity. Until now, the method described above is called SABA, which means a single-layer ABA.

As we can see, SABA is inherently a local optimization based method, which detects communities by only using individual vertex's movements among communities. Thus, it can easily obtain the community structure with a high resolution, while this attained result may not be a partition corresponding to the maximum modularity *Q*.

In order to further improve the performance of our algorithm, a strategy of "layer and rule" is introduced into it, and this extends the single-layer ant-based algorithm SABA to a multi-layer ant-based algorithm MABA. In MABA, we firstly run SABA on the original network as the first level, so as to attain a high resolution community structure of this network. Then, we will reconstruct a higher level networks based on this partition, which makes each detected community as a vertex and the sum of the weights of edges between any two communities as the weight between the aforementioned vertices. Thereafter, we will run SABA again on the new generated network at this higher level. To sum up, the process of MABA can be described as follows: we iteratively run SABA on the network at the current level and reconstruct a higher level network for the next run of SABA, and finally stop it when there is no increase of modularity *Q* whatsoever. At last, we select the partition corresponding to the maximum *Q*-value from the hierarchical community structures and treat it as the best one.

As we can see, from the microcosmic angle, the single-layer algorithm SABA is to optimize the global function *Q* by making each ant optimize the local function *f* of its current vertex, with the aid of a simulated annealing idea. From the macroscopic perspective, SABA also can be seen as a label diffusion process, which is propagated by each of the ants from one vertex to another. In addition, our multi-layer algorithm MABA can further improve the single-layer SABA by implementing merging and splitting communities, which is accomplished through the vertices' movements among communities on the higher level network.

### 3.3. *Single-layer ant-based algorithm with local optimization*

Based on the above discussions, here we give a complete and detailed description of the single-layer ant-based algorithm with local optimization (SABA), which is shown as Fig. 1.

In SABA, we firstly initialize each vertex with a unique label. So, in this initial partition there are as many communities as there are vertices. Thereafter, the $n'$ ants are randomly distributed to the vertices of the input network. Here, $n' = p \times n$, where *n* is the number of vertices and *p* denotes the proportion covered by the population of the ant colony in the number of vertices.

Then the algorithm proceeds in a number of iterations. In each iteration, each ant follows the same set of actions, which consists of 'ant moves' and 'ant optimizes'. For the first action 'ant moves', the ant randomly selects one of its neighbors with a different label if possible, and moves to this selected vertex. Otherwise, if the ant and all its neighbors have a same label, then it randomly selects one of its neighbors and moves to it (but need not carry on the 'ant optimizes' action). For the second action 'ant optimizes', the ant decides whether to accept its previous vertex's label as its current vertex's new label with the aid of a simulated annealing strategy, whose purpose is to optimize the global function *Q* from its local perspective. This strategy is that the ant accepts the label of its previous vertex with an annealing probability:

$$Pro = \begin{cases} 1 & \text{if } f'_{cur} > f_{cur} \\ \exp(-\dfrac{f_{cur} - f'_{cur}}{T}) & \text{if } f'_{cur} < f_{cur} \end{cases}, \quad (3)$$

where $f_{cur}$ is the *f*-value of the current vertex when it is assigned to its own label, $f'_{cur}$ is the *f*-value of this current vertex when it accepts the label of its previous vertex, and *T* is annealing temperature. Note that the calculation of function *f* for each node is shown as Eq. (2). After each iteration, this temperature is cooled down by $T = c_T \times T$, where $c_T$ is the annealing coefficient.

Ideally, the iterative process of SABA should continue until no vertex in the network changes its label by the ants. However, there may be vertices that have an equal maximum value of function *f* in two or more communities. For each of these vertices, its label could still change over iterations even if the labels of all related vertices (vertices in communities containing at least one neighbor of the current vertex) remain constant. Hence we perform the iterative process until no further improvement on modularity can be achieved, i.e., until a local maxima of the modularity is attained. The experimental result shows that, the convergence property of our SABA is very good. Even for large-scale networks containing millions of edges, it can still converge within just 50 iterations.

It seems that, SABA looks like a label passing process with local optimization, but from the algorithm level it is in fact a type of ant-based technique. Ant-based algorithms are optimization techniques that imitate the collective ability of an ant colony to solve problems [27]. In SABA, a colony of ants is randomly produced, and each of them follows the same set of rules. These ants (or called mobile-agents) interact with each other by the 'blackboard' architecture which is a type of agent communication mechanism, and all of them work collectively so as to detect the communities of the network. It is obvious that our SABA possesses the nature of ant-based algorithms. Thus, although SABA makes use of the techniques of label passing and local optimization, it still belongs to ant-based algorithms.

Moreover, SABA also has some distinct differences compared with the existing ant-based algorithms for community detection. Unlike most of those methods where ants correspond with each other through pheromone, our ants (or called mobile-agents) communicate with one another by a particular underlying interactive mechanism which makes the actions of the current ants affected by that of all the previous ones. In other words, all our ants work collectively on a same community division, which helps to realize the indirect communication among the ant colony. This special interactive mechanism, which does not employ pheromone, significantly reduces the running time of SABA and makes it well suitable for large-scale networks. Meanwhile, the succeeding experiments show that it also performs very well in terms of clustering quality.

To sum up, the process of SABA makes each vertex optimize its own function *f*, which finally achieves the purpose to optimize the modularity *Q* of the entire network. Meanwhile, SABA can also be regarded as a process where vertices are constantly moved among communities so as to detect the underlying community structure of the network.

Procedure   SABA

Input:   N, p, T, $c_T$; // N denotes the network, p is the proportion of ant colony, T is the initial temperature, and $c_T$ is the annealing coefficient

Output:   CP; // the community partition

Begin

1  $l \leftarrow 0$; // l is the iteration number

2  For $\forall v \in V$, $C_v(l) \leftarrow v$; // initialize the labels of all vertices. V denotes the set of vertices and C(l) denotes the partition of the l-th generation

3  $Q(l) \leftarrow$ Compute the value of modularity for $C(l)$; // Q(l) denotes the Q-value of C(l)

4  Randomly distribute $n'$ ants on network N; // $n' = p*n$, n is the number of vertices

5  Do

6      $l \leftarrow l+1$;

7      $C(l) \leftarrow C(l-1)$;

8      For $j = 1: n'$

9          If the vertex of ant j and all its neighbors have the same label

10             Ant j randomly selects one of its neighbors and moves there;

11         Else

12             previous_vertex ← The vertex where ant j is situated;

13             current_vertex ← Randomly selects one of its neighbors with a different label;

14             Ant j moves to current_vertex;

15             $f_{cur} \leftarrow$ Compute the f-value of current_vertex with its own label;

16             $f'_{cur} \leftarrow$ Compute the f-value of current_vertex with the previous_vertex's label;

17             If $f'_{cur} > f_{cur}$

18                 $C_{current\_vertex}(l) \leftarrow C_{previous\_vertex}(l)$ with probability 1;

19             Else

20                 $C_{current\_vertex}(l) \leftarrow C_{previous\_vertex}(l)$ with the annealing probability P;

                   // $P = \exp\left(-\dfrac{f_{cur} - f'_{cur}}{T}\right)$, a type of simulated annealing strategy

21         End

22      End

23  End

24  $T \leftarrow T*c_T$; // cooling

25  $Q(l) \leftarrow$ Compute the value of the modularity for $C(l)$;

26 Until   $abs(Q(l)-Q(l-1))<1.0e-6$

27 $CP \leftarrow C(l)$;

End

**Fig. 1.** The algorithm flow of SABA

**Proposition 1.** The time complexity of SABA cannot be worse than $O(c \times n)$, where $n$ is the number of vertices and $c$ is the average community size in the end of this algorithm.

*Proof.* It's obvious that, the steps 11 and 12 are most computationally expensive in SABA, and the time complexity of these two steps depends on the size of the community which the current vertex belongs to. As the scale of each community is always increasing with time passing in the SABA, the average community size at any iteration should be smaller than the average community size $c$ at the end of the algorithm. Therefore, the running time of these two steps can not be greater than $O(c)$ at one time. Furthermore, each of these two steps will be carried out $n'$ ($n' = p \times n$) times at each iteration, and there are $l$ iterations in this algorithm. Thus, the time complexity of the SABA will not be greater than $O(l \times p \times n \times c)$. Furthermore, as parameters $l$ and $p$ can both be regarded as a constant, the complexity of our method can be also given by $O(c \times n)$ now.

### 3.4. Multi-layer ant-based algorithm with local optimization

Obviously, SABA is inherently a local optimization based method, and it detects communities just by making use of single vertices' movements among communities. Although it can easily attain the community structure with a high resolution, this result may not correspond to a partition with the maximum $Q$-value. Therefore, here we present a multi-layer algorithm MABA to further improve the single-layer algorithm SABA by adding two types of actions (i.e. merging and splitting communities), which are implemented through vertices' movement on the networks at a higher level. The description of MABA is in Fig. 2.

Procedure    MABA

Input:    $N$; // *the original network*

Output:    *best_Partition*; // *the partition corresponding to the maximal Q-value*

Begin

1 $i \leftarrow 0$;

2 $N(i) \leftarrow N$; // *N(i) denotes the i-level network*

3 Do

4    $H(i) \leftarrow$ Run SABA on $N(i)$; // *H(i) denotes the partition of N(i)*

5    $Q(i) \leftarrow$ Compute the $Q$-value of $H(i)$; // *Q(i) denotes the Q-value of H(i)*

6    $i \leftarrow i+1$;

7    $N(i) \leftarrow$ Construct a higher level network based on $H(i)$; // *take each community as a vertex, and the sum of the weight of edges between any two communities as the weight between the new vertices*

8 Untill    $Q(i) \leq Q(i-1)$

9 *best_Partition* $\leftarrow H(i-1)$; // *the partition with the maximal Q-value*

End

**Fig. 2.** The algorithm flow of MABA

As we can see, the process of network reconstruction in step 7 will produce a new weighted network at a higher level, which has self-loop edges. While, fortunately, SABA has the ability to deal with this type of networks. Meanwhile, function $Q$ is also suitable for weighted network with self-loop edges, and the $Q$-value of the community division of a high level network indeed is equal to the same quantity for the community division obtained by mapping this result to its original network. In fact, running SABA on different level of networks can all be regarded as the optimization for modularity $Q$ of the original network.

Based on the strategy of "layer and rule", in our multi-layer algorithm MABA, we firstly make the single-layer method SABA unfold the communities from the network at the current level; and then, we reconstruct a higher level networks based on this partition (takes each detected community as a vertex and the sum of the weights of edges between any two communities as the weight between the aforementioned vertices), which is prepared for the next run of SABA at a higher level. This is an iterative process. Obviously, MABA naturally incorporates a notion of hierarchy as communities of communities are built during the process, thus it can easily discover hierarchical community structures that are intrinsic in most real networks. However, the tools to analyze hierarchy are not as advanced as the tools for communities, and that the full structure is currently more difficult to handle than communities. Thereby, here we still select the partition corresponding to the maximal $Q$-value as a final result from the attained hierarchical community structures by step 9.

As we can see, our single-layer algorithm SABA detects community structure only by moving vertices between communities. While, our multi-layer method MABA also facilitates merging two communities and splitting a single community, which are implemented by moving vertices on higher level networks. According to Ref. [13], combining these three operations (moving vertex, merging and splitting communities) can make a community detection algorithm have stronger ability for searching. Thus, MABA has more powerful search ability and can attain a community division with a higher $Q$-value. Moreover, MABA has some distinct differences compared with the conventional hierarchical algorithms. Specifically, agglomerative hierarchical methods (e.g. FN) can only merge communities, while divisive hierarchical methods (e.g. GN) can only split communities. But our algorithm can iteratively agglomerate and split the communities achieved through moving vertices on networks at a higher level. Thus it has stronger search ability and can obtain a better solution.

Modularity $Q$ as a important metric has been widely accepted by the scientific community. But Fortunato and Barthélemy [6] recently proved that the modularity optimization methods often suffer from the resolution limit problem and fail to discover the communities with less than $\sqrt{L/2}$ edges, where $L$ is the number of edges in the network. Fortunately, our method MABA can mitigate the resolution limit of modularity due to its intrinsic multi-level nature. To be specific, at the first level, MABA only employs the displacement of single nodes from one community to another. Obviously, the probability that two distinct communities can be merged by moving vertices one by one is very low. Consequently, a quite high resolution community division will be attained in the end of the first level of our method. Experimental results on several large-scale real networks also show that, at the first level of MABA, more than 95% of detected communities have fewer than $\sqrt{L/2}$ edges. With the increase of the level number of this method, these communities may possibly be merged at higher levels, after blocks of vertices have been aggregated. Then some partitions with relatively lower resolutions may be obtained. At last, our algorithm can provide a decomposition of the network into communities for different levels of organization, thereby giving access to different resolutions of community detection. Moreover, the intermediate solutions found by our algorithm may also be meaningful, and the uncovered hierarchical structure may allow the end-user to zoom in the network and to observe its structure with the desired resolution. What is more, some algorithms, which provide multi-scale community structures by adjusting resolution parameter, may artificially force the disaggregation of large and densely connected communities. Instead, MABA can naturally attain different resolution partitions.

**Proposition 2.** The time complexity of MABA is O($c \times n$) in the worst case.

***Proof.*** From the analysis of proposition 1, the time complexity of its SABA subroutine can not be worse than O($c \times n$). The MABA will iteratively run its sub-method SABA on networks at different levels for several times. While, fortunately, the network scale will decrease very fast with the increase of the level number, and the maximal level number is also very small in general from our experiment. Thus, the most computationally expensive step in our MABA is that, the SABA subroutine is carried out on the original network at the first level. Therefore, the worst-case time complexity of the MABA can be given by O($c \times n$).

It's worth mentioning that, the sizes of well-defined communities ($c$) in large-scale real networks are always much smaller than the scales of networks ($n$) [28]. Thus, MABA will be very efficient, and even takes an almost linear time to deal with the real large networks.

## 4. Experiments

In order to evaluate the performance of MABA, we tested it on two types of benchmark artificial networks, as well as in some widely used large-scale real networks. Then we provide an analysis to illustrate that MABA can mitigate the resolution limit of modularity. Thereafter, we also offer a simple example to validate its potential capability to uncover multi-scale hierarchical structures. At last, we conclude by analyzing the convergence property of MABA, which is especially important for this algorithm.

There are three parameters: $T$, $c_T$ and $p$ in MABA. Both $T$ and $c_T$ are simulated annealing parameters. $T$ denotes the initial temperature, and $c_T$ denotes annealing coefficient. While, $p$ is a parameter of ant-base algorithm, which denotes the fraction of ant colony size compared with the number of vertices in the network. According to our experience and some experiment results, we set $T = 500$, $c_T = 0.1$ and $p = 0.6$ in this paper.

In the experiments, MABA is compared with five representative as well as efficient community detection algorithms, which are FN [12], FEC [17], Infomap [18], FUA [8] and RN [19]. It is noted that, algorithms FUA, Infomap and RN are considered to have an excellent performance by the famous survey [11]. All experiments are done on a single Dell Server (Intel(R) Xeon(R) CPU 5130 @ 2.00GHz 2.00GHz processor with 4Gbytes of main memory), and the source code of the algorithms used here can all be obtainable from the authors.

### 4.1. Computer-generated networks

Here we evaluate the performance of MABA on two kinds of widely used benchmarks with built-in community structure. They are the Girvan and Newman (GN) benchmark [3] and the Lancichinetti-Fortunato-Radicchi (LFR) benchmark [29], respectively. To estimate the goodness of the answer given by the algorithm as compared to the real answer that is expected, here we employ a widely used accuracy measure so called Normalized Mutual Information (NMI) [4]. The NMI measure which makes use of information theory is regarded as a comparative fair metric compared with the other ones [4].

#### 4.1.1. The GN benchmark

The first type of synthetic networks employed here is proposed by Girvan and Newman [3]. For this benchmark, each graph consists of $n = 128$ vertices divided into 4 groups of 32 vertices. Each vertex has on average $z_{in}$ edges connecting it to members of the same group and $z_{out}$ edges to members of other groups, with $z_{in}$ and $z_{out}$ chosen such that the total expected degree $z_{in}+z_{out} = 16$, in this case. As $z_{out}$ is increased from the small initial values, the resulting graphs pose greater and greater challenges to the community detection algorithms. In Fig. 3(a), we show that the NMI accuracy attained by each algorithm is as a function of $z_{out}$ from 1 to 12. As we can see, our algorithm MABA outperforms all the other five methods in terms of NMI accuracy on this benchmark.

Computing speed is another very important criterion to evaluate the performance of an algorithm. Time complexity analysis for MABA has been given by Proposition 2 in Sec. 3.4. Nevertheless, here we also show the actual running time of MABA from an experimental angle in order to further evaluate its efficiency. For this application, each graph consists of $n = 100K$ vertices divided into $K$ groups of 100 vertices. Each vertex has on average $z_{in} = 10$ edges connecting it to members of the same group and $z_{out} = 6$ edges to members of other groups. The only difference between the networks used here and the former ones is that, now $z_{out}$ is fixed while the communities number $K$ is changeable. Fig. 3(b) shows the experimental result. As we can see, the actual running time of MABA is proportional to the scales of networks, under the condition that the average community size of actual network community structure is a constant. Thus, this experiment can also validate the correctness of Proposition 2, which depicts the time complexity of MABA as $O(c \times n)$.

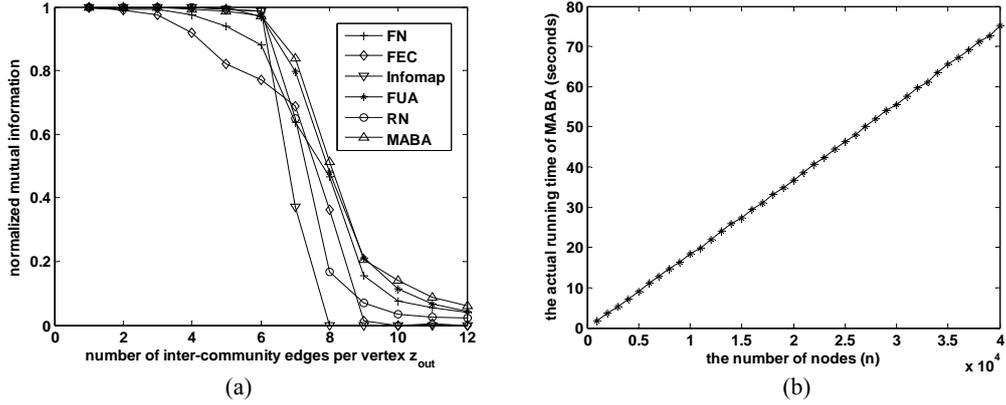

**Fig. 3.** Test the performance of MABA on the GN benchmarks. (a) Compare MABA with FN, FEC, Infomap, FUA and RN in terms of NMI accuracy. (b) The actual running time of MABA as a function of the network scale. Each point is an average over 50 realizations of the graphs. Lines between points are included solely as a guide to the eye.

### 4.1.2. *The LFR benchmark*

In order to further evaluate the accuracy of these algorithms, a new type of benchmark proposed by Lancichinetti et al. [29] is also adopted here. Unlike the GN benchmark where all the vertices have a unified degree and all the community sizes are the same, both the degree and the community size distributions in the LFR benchmark are power law, which is a statistical property that most real-world networks seem to share.

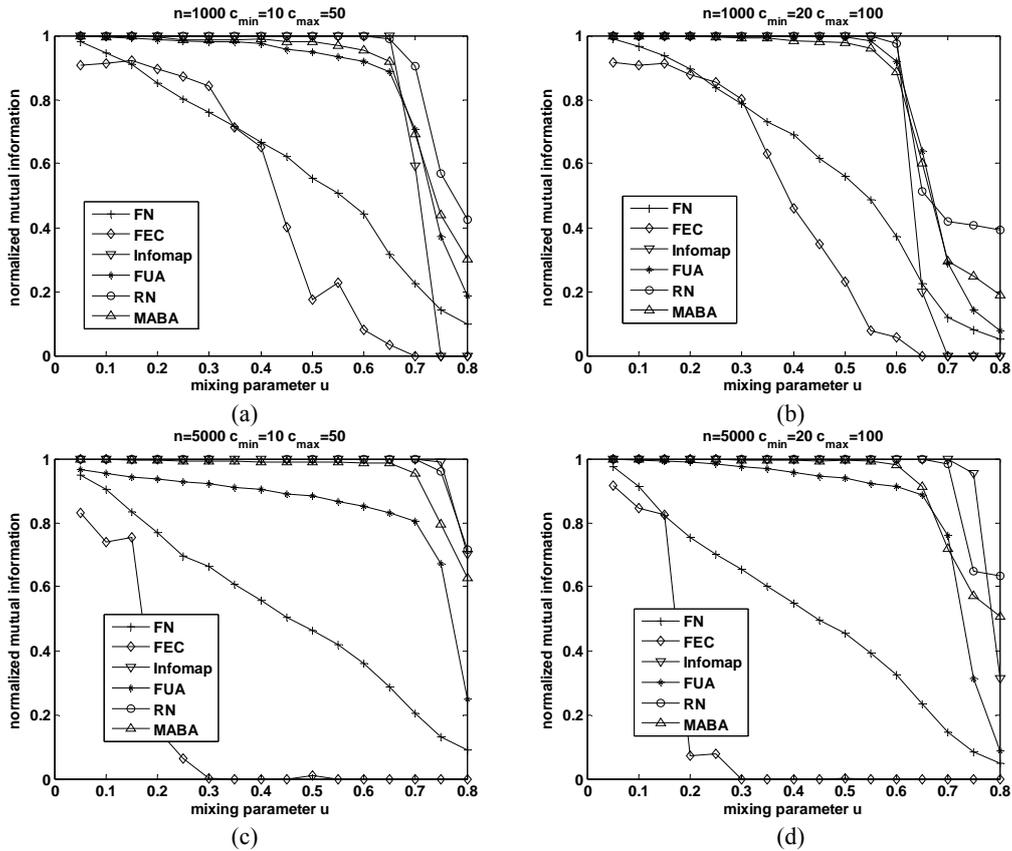

**Fig. 4.** Compare MABA with FN, FEC, Infomap, FUA and RN in terms of NMI accuracy on the LFR benchmark networks. Each point is an average over 50 realizations of the graphs. Lines between points are included solely as a guide to the eye. (a) Comparison on small networks with small communities ($n = 1000$, $c_{min} = 10$, $c_{max} = 50$). (b) Comparison on small networks with big communities ($n = 1000$, $c_{min} = 20$, $c_{max} = 100$). (c) Comparison on big networks with small communities ($n = 5000$, $c_{min} = 10$, $c_{max} = 50$). (d) Comparison on big networks with big communities ($n = 5000$, $c_{min} = 20$, $c_{max} = 100$).

Following the experiment designed by survey [11], the parameters setting for the LFR benchmark networks are as below. The network size $n$ is set to either 1000 or 5000, the minimum community size $c_{min}$ is set to either 10 or 20, and the mixing parameter $\mu$ (each vertex shares a fraction $\mu$ of its edges with vertices in other communities) varies from 0 to 0.8 with interval 0.05. We keep the remaining parameters fixed: the average degree $d$ is 20, the maximum degree $d_{max}$ is $2.5\times d$, the maximum community size $c_{max}$ is $5\times c_{min}$, and the exponents of the power-law distribution of vertex degrees $\tau_1$ and community sizes $\tau_2$ are -2 and -1 respectively. In Fig. 4, we show that the NMI accuracy attained by each algorithm is as a function of the mixing parameter $\mu$. As we can see, MABA is close to Infomap and RN, and outperforms the other three methods in terms of NMI accuracy on this more challenging benchmark.

### 4.2. Real-world networks

As real networks may have some different topological properties from the synthetic ones, here we adopt several widely used large-scale real networks to further evaluate the performance of these algorithms. These networks that we used here are all listed in Table 1. The sizes of these networks range from thousands of vertices to (near) millions of vertices.

Table 1. Real-world networks used here.

| Networks | $|V|$ | $|E|$ | Descriptions |
|---|---|---|---|
| word | 7,207 | 31,784 | Semantic network [30] |
| internet | 22,963 | 48,436 | A snapshot of the Internet by Mark Newman [31] |
| arxiv | 56,276 | 315,921 | Scientific collaboration networks [32] |
| www | 325,729 | 1,090,108 | Edgeed WWW pages in the nd.edu domain [33] |
| amazon | 473,315 | 3,505,519 | Amazon products from 2003 all [34] |
| webgoogle | 855,802 | 4,291,352 | Web graph Google released in 2002 [35] |

Because the inherent community structure for real networks is usually unknown, here we adopt the most commonly used modularity $Q$ [5] to evaluate the performance of these algorithms. Table 2 shows the average result (over 50 runs) that compares our method MABA with FN, FEC, Infomap, FUA and RN in terms of function $Q$ on the real-world networks described in Table 1. The empty cells correspond to computation time over 24 hours or out of memory. As we can see, the clustering quality of MABA is competitive with that of FUA, and better than that of the other four algorithms. This also shows that MABA is very effective on large-scale real networks.

Table 2. Compare MABA with FN, FEC, Infomap, FUA and RN in terms of function $Q$ on large-scale real networks.

| $Q$-value | FN | FEC | Infomap | FUA | RN | MABA |
|---|---|---|---|---|---|---|
| word | 0.4665 | 0.4609 | 0.4617 | 0.5246 | 0.2205 | 0.5150 |
| internet | 0.6378 | 0.6104 | 0.5783 | 0.6613 | 0.0875 | 0.6487 |
| arxiv | 0.7153 | 0.7276 | 0.6861 | 0.7801 | - | 0.7724 |
| www | - | 0.7962 | 0.8832 | 0.9455 | - | 0.9315 |
| amazon | - | 0.8088 | 0.7272 | 0.8478 | - | 0.8473 |
| webgoogle | - | 0.9409 | 0.8535 | 0.9771 | - | 0.9719 |

### 4.3. Resolution limit analysis

In order to illustrate that MABA can mitigate the resolution limit of modularity, we first test it on two classical artificial networks designed by Fortunato and Barthélemy [6], and then we also apply it on several large-scale real networks used in Sec. 4.2.

#### 4.3.1. Classical synthetic networks

Here we will test our method on the synthetic networks by which Fortunato and Barthélemy demonstrated the resolution limit of modularity [6].

The first test case is shown as Fig. 5(a). It is a network with two pairs of identical cliques: the big pair has 20 vertices each, while the other one is a pair of 5-vertex cliques. This network has a clear modular structure where each community corresponds to a single clique, thus a proper community detection algorithm should be able to find them correctly. But modularity optimization methods often combine the pair of 5-vertex cliques into one community. This is because the maximal modularity of this network corresponds to a partition where these two small cliques are merged together. However, the first level of our algorithm finds the natural partition of the network, where each community corresponds to one unique clique. The $Q$-value of this partition is 0.5416. The second level of this method finds the partition corresponding to the global maximum of modularity where the two smaller cliques are merged (as shown by the ellipse in Fig. 5(a)). The $Q$-value of this partition is 0.5426 which is larger than the former one. As we can see, the cliques are indeed merged in the higher level partition due to the resolution limit, but they are still distinct in the first level partition. Thus, MABA can not only provide a high resolution division, but also can offer the partition corresponding to the maximal modularity.

The second test case is shown as Fig. 5(b). It consists of a ring of cliques, connected through single links. There are totally 30 cliques, and each one contains 5 nodes. Again, the first level of our algorithm discovers the natural partition of this network, where each community corresponds to one 5-vertex clique. The third level of this method finds the solution with maximum modularity, where cliques are combined into groups of 2. The result also indicates that our method seems able to mitigate the resolution limit.

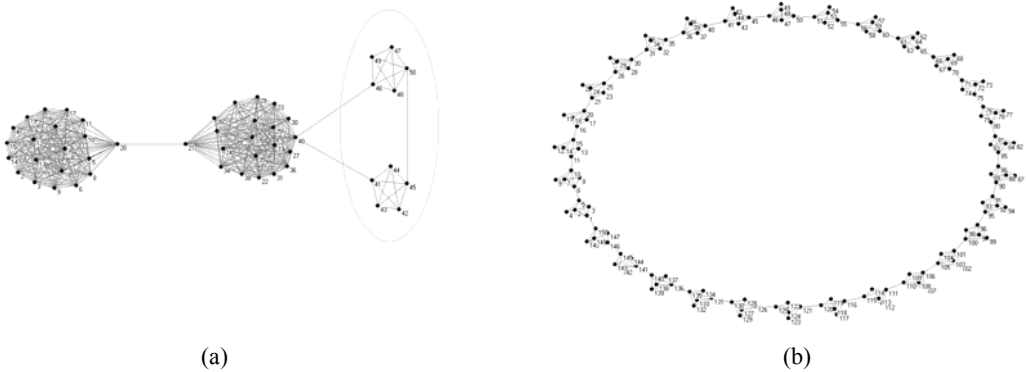

(a)            (b)

**Fig. 5.** Illustration of two types of clique chain networks which are usually used to demonstrate the problem of the resolution limit with modularity. (a) Network made of two pairs of identical cliques, with big pair having 20 vertices each and the other pair having 5 vertices each. (b) Network made of 30 cliques, with 5 vertices each, connected by a single edge.

We further performed testing on ten similar clique chain networks by varying the number of cliques. Here, the clique size is fixed at 5, while the clique number ranges from 10 to 100 with interval 10. Considering the resolution limit problem, with the increase of clique number, the resulting networks will pose greater and greater challenges to the modularity optimization algorithms. Besides MABA, here we also apply two pure modularity optimization approaches (i.e., FN and SA) on these networks as contrast experiments. FN is a widely used greedy modularity optimization method, and SA is one of the most accurate modularity maximization algorithms. The experimental results are summarized in Table 3. As we can see, for networks containing 10 or 20 cliques with sizes larger than $\sqrt{L/2}$, all the three algorithms can detect the actual number of cliques correctly. For networks containing more than 30 cliques, algorithms FN and SA both fail to uncover the actual partitions. This is because the clique size is less than $\sqrt{L/2}$ for each of these networks, which makes most modularity optimization methods suffer from the resolution limit problem. However, MABA can still effortlessly find the real communities at its first level. Thus, this can prove that MABA is able to mitigate the resolution limit of modularity.

Table 3. The resolution limit test on networks made of identical cliques connected by single links.

| Methods | Number of cliques | | | | | | | | | |
|---|---|---|---|---|---|---|---|---|---|---|
| | 10 | 20 | 30 | 40 | 50 | 60 | 70 | 80 | 90 | 100 |
| FN | 10 | 20 | 15 | 20 | 25 | 30 | 35 | 40 | 23 | 25 |
| SA | 10 | 20 | 16 | 22 | 20 | 23 | 32 | 35 | 29 | 36 |
| MABA | 10 | 20 | 30 | 40 | 50 | 60 | 70 | 80 | 90 | 100 |

4.3.2. *Real-world networks*

To further illustrate that MABA is able to mitigate the resolution limit, here we also provide some detailed analysis on the aforementioned six real networks in terms of several metrics. An introduction to these metrics is given as follows. $C$ is the number of communities. $Q$ is the well-known modularity function. $C_1$ is the number of communities with fewer than $\sqrt{L/2}$ edges, and $P_1$ ($P_1=C_1/C$) denotes the fraction of these communities. $C_2$ is the number of communities with edge number between $\sqrt{L/2}$ and $\sqrt{2L}$, and $P_2$ ($P_2=C_2/C$) denotes the fraction of this type of communities. $C_3$ is the number of communities which meets the "weak" definition given by Radicchi et al. [36], and $P_3$ ($P_3=C_3/C$) is the fraction of this kind of communities. $D$ is a partition density function which is based on edge density inside communities. Unlike modularity $Q$, $D$ does not suffer from a resolution limit [37]. We run MABA on each of these networks, and obtain its first level partition and the partition corresponding to the maximal modularity. The community results of these two levels in terms of the above metrics are summarized as Table 4.

Table 4. Compares the high resolution partition with the one at the maximal modularity for several real networks.

| Networks | Level | $C$ | $Q$ | $C_1$ ($P_1$) | $C_2$ ($P_2$) | $C_3$ ($P_3$) | $D$ |
|---|---|---|---|---|---|---|---|
| word | The first | 600 | 0.4167 | 583 (97.1667%) | 7 (1.1667%) | 67 (11.1667%) | 0.0939 |
| | Maximal $Q$ | 25 | 0.5204 | 4 (16%) | 1 (4%) | 22 (88%) | 0.0094 |
| internet | The first | 2,036 | 0.5013 | 2,008 (98.6248%) | 8 (0.3929%) | 1005 (49.3615%) | 0.0051 |
| | Maximal $Q$ | 136 | 0.6532 | 107 (78.6765%) | 11 (8.0882%) | 135 (99.2647%) | 0.0025 |
| arxiv | The first | 5,317 | 0.6403 | 5,170 (97.2353%) | 47 (0.8840%) | 3,402 (63.9835%) | 0.2726 |
| | Maximal $Q$ | 327 | 0.7725 | 260 (79.5107%) | 20 (6.1162%) | 324 (99.0826%) | 0.0235 |
| www | The first | 12,743 | 0.8555 | 12,305 (96.5628%) | 91 (0.7141%) | 11,298 (88.6604%) | 0.0561 |
| | Maximal $Q$ | 353 | 0.9330 | 193 (54.6742%) | 34 (9.6317%) | 353 (100%) | 0.0060 |
| amazon | The first | 21,353 | 0.6766 | 20,498 (95.9959%) | 98 (0.4590%) | 13946 (65.3117%) | 0.2300 |
| | Maximal $Q$ | 731 | 0.8486 | 501 (68.5363%) | 74 (10.1231%) | 675 (92.3393%) | 0.0128 |
| webgoogle | The first | 48,983 | 0.7916 | 47,631 (97.2399%) | 157 (0.3205%) | 41,677 (85.0846%) | 0.1863 |
| | Maximal $Q$ | 2,199 | 0.9697 | 1,975 (89.8136%) | 64 (2.9104%) | 2,152 (97.8627%) | 0.0096 |

As we can see, for each of these networks, the number of communities in the first level partition is far larger than the number of modules obtained at the maximum modularity. Meanwhile, we find that the attained community structure at the first level also corresponds to a very high value of $Q$, even though it is a little smaller than the value of maximum modularity.

It is proved that modularity optimization may fail to discover the communities with less than $\sqrt{L/2}$ edges [6]. This is called a resolution limit problem. For each of these networks, the first level partition of MABA has more than 95% of modules with fewer than $\sqrt{L/2}$ edges. This result clearly shows that MABA is not restricted by this resolution limit problem. We further investigate whether these communities at the first level make some sense in terms of a well-known metric namely "weak" definition. For each of these networks, most of the communities in the first level partition can satisfy the "weak" definition. Thus, it shows that this high resolution partition found by our algorithm has a well-defined community property. There is just one exception, which is the "word network". For this network, the fraction of communities which meet the "weak" definition is relatively low. This may be because that it contains lots of highly overlapping communities, which can have many more external than internal connections [37].

Here, we also offer some comparisons between the first level partition and the partition at the maximum modularity in our algorithm. As we can see, for each of these networks, the fraction of communities with fewer than $\sqrt{L/2}$ edges at the first level is far larger than the same quantity at the maximum modularity. According to Ref [6], this type of communities may not be found by modularity optimization because of the resolution limit problem. Furthermore, the community number which meets the "weak" definition in the first level partition is even far larger than the total number of communities in the partition corresponding to maximum modularity. This indicates that most of the communities which meet the "weak" definition at the first level may be merged in the maximal modularity partition. Moreover, the fraction of communities with edge number between $\sqrt{L/2}$ and $\sqrt{2L}$ at the first level is much smaller than the same quantity at the maximum modularity. Known from [6], the probability that this type of modules conceals substructures is relatively large, which may lead to the resolution limit problem. From the above three points, we can conclude that the first level of our method can easily offer the high resolution community structures, which may mitigate the resolution limit very well.

Since the modularity $Q$ processes the resolution limit problem, here we introduce another type of measure for the quality of a community partition, which is claimed to be free from this problem [37]. This new metric is called "partition density $D$", and defined for link communities [37]. In order to make it fit for our situation, here we generalize it to vertex communities and define a "vertex partition density" as follows.

For a network with $n$ vertices, $P = \{P_1, P_2, \cdots, P_S\}$ denotes a partition of the vertices into $S$ subsets. Subset $P_s$ has $n_s = |P_s|$ vertices and $m_s = |\cup_{i \in P_s, j \in P_s} e_{ij}|$ edges. Then we define the density of communities $P_s$ as Eq. (4).

$$D_s = \frac{m_s - (n_s - 1)}{n_s(n_s - 1)/2 - (n_s - 1)} \tag{4}$$

This is $m_s$ normalized by the minimum and maximum numbers of vertices possible between $n_s$ connected vertices (We assume that $D_s=0$ if $n_s=2$). The "vertices partition density" $D$, is the average of $D_s$ weighted by the fraction of present vertices:

$$D = \frac{2}{n} \sum_s \frac{n_s(m_s - n_s + 1)}{(n_s - 2)(n_s - 1)} \tag{5}$$

According to Table 4, for each of these networks, the "vertices partition density" of our first level partition is indeed far larger than the same quantity of the maximal modularity partition. This means that the community property of the high resolution partition obtained by the first level of our algorithm is very well, it's even better than the partition corresponding to the maximum modularity. This can further proves that our resolution limit free solution also has a very high community quality.

**4.4. *Multi-scale hierarchy***

The key point of our algorithm MABA is to attain the community division with a high clustering quality. However, this method also possesses the naturally ability to uncover the hierarchical structures with a different granularity. Thus we offer a simple example here so as to show the potential ability of MABA to unfold hierarchy by its multi-level mechanism.

We adopt the US college football association network [3] as a sample graph in this example. It contains 115 vertices and 613 edges, which correspond to football teams and games played among teams, respectively. All teams are divided into 12 conferences. Each conference is considered as one actual community since the number of games played within the same conference should be much more than those between conferences.

We randomly run MABA once for the football network. The result is shown as Fig. 6. As we can see, for the network at the first level, our method attains a community division containing 10 well-defined communities, which is shown as Fig. 6(a). Its NMI accuracy is 0.8923, while the clustering quality ($Q$) is 0.6044, which is even much higher than that of the actual community structure ($Q = 0.5518$). At the second level, it attains a community division with a relatively lower resolution, which contains 6 well-defined communities. This is shown as Fig. 6(b). Its NMI accuracy is 0.7448, while the clustering quality ($Q$) is 0.5817, which is still a little higher than that of the actual community structure ($Q = 0.5518$). At the third level, which is also the last level, it attains a community division with the lowest resolution which just contains 2 well-defined communities, and its $Q$-value is 0.3846. This is shown as Fig. 6(c). According to the analysis of Ref. [5], the community structure of a network is obvious when its $Q$-value is greater than 0.3. Thus, this community division with the lowest resolution should still have some valuable senses, even though it just contains two communities. Therefore, we can say that our method also has the potential ability to effectively detect the reasonable hierarchical community structures, which are found to be intrinsic in most real networks [38].

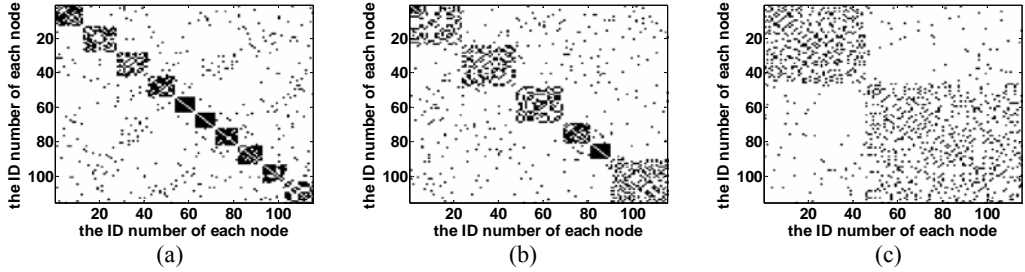

**Fig. 6.** The multi-scale hierarchical structures got by MABA on the football network. We describe the community structure at each level as a diagonal matrix, which is sorted according to this partition. (a) The network clustering solution at the first level. The corresponding $Q$-value is 0.6044, community number is 10 and NMI accuracy is 0.8923. (b) The network clustering solution at the second level. The corresponding $Q$-value is 0.5817, community number is 6 and NMI accuracy is 0.7448. (c) The network clustering solution at the third level. The corresponding $Q$-value is 0.3846, community number is 2 and NMI accuracy is 0.3515.

### 4.5. *Convergence analysis*

As follows, we discuss the convergence property of MABA. We firstly give the convergence analysis of its SABA subroutine with the increase of the iteration number on the original network at the first level. Thereafter, we will show that the $Q$-value got by MABA is as a function of the level number.

In MABA, it's obvious that the original network at the first level is the largest. Thus, if its SABA subroutine can converge on the network at the first level, it will be necessary convergent on the higher level networks. Fig. 7(a) shows the convergence process of SABA with the increase of iteration number at the first level of each real network. As we can see, against the networks containing thousands or even (near) millions of vertices, the SABA subroutine can still converge very well within 50 iterations.

MABA will stop when the $Q$-value got by its SABA subroutine at the current level is lower than that at the previous level, and then returns the maximum $Q$-value. Thus, it's very important that we know the level number which corresponds to the maximal $Q$-value. Fig. 7(b) shows that the $Q$-value got by MABA is as a function of the level number on each real network. As we can see, against each of these networks, the $Q$-value which corresponds to the second level is the highest. Thus, it's obvious that MABA can stop at the third level in general.

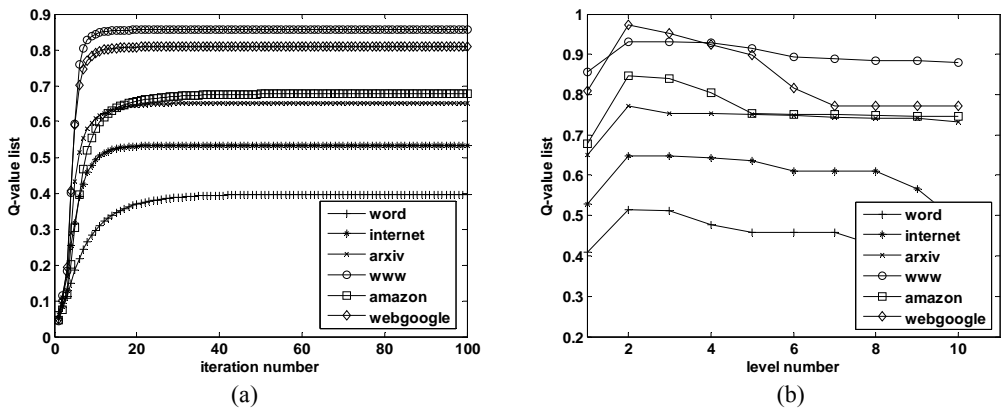

**Fig. 7.** The convergence analysis of MABA on all the real networks used here. (a) The convergence property of its SABA subroutine with the increase of the iteration number on each network at the first level. (b) The $Q$-value got by MABA as a function of the level number.

### 5. Discussion and Conclusions

The main contribution of this paper is that, a multi-layer ant-based algorithm with a type of local optimization strategy, which can discover high-quality community structure with low computational complexity, has been proposed. At each single level, this algorithm is indeed a local optimization based method, in which each ant tries to propagate its label to one of its neighbors under the control of a simulated annealing idea. While, from the multi-level angle, it is actually a process which locally optimizes modularity $Q$ through three types of operations: displacement of vertices, merging two communities and splitting a community. Especially, the actions of merging and splitting communities are implemented by moving vertices on higher level networks. Known from Ref. [13], this process (iteratively moving vertices, merging two communities and splitting a community) is very effective to produce well candidate solutions. Thus it makes our algorithm able to attain the community division with a high clustering quality. This has been verified by a set of rigorous experiments which compares our MABA with several representative algorithms on both computer-generated benchmarks and some widely used real-world networks. Meanwhile, the time complexity analysis and experiments about the actual running time of our MABA demonstrate that, this method is quite efficient, and even a near linear algorithm for large-scale real networks. Last but not least, MABA can mitigate the resolution limit problem of pure modularity optimization, which is also validated by a set of experiments.

In this work, we mainly focus on modularity optimization and its associated resolution limits, and the problem of extreme degeneracy is not considered here. However, as we know, Good et al. [7] pointed out that, "only if the network is relatively small and contains a few non-hierarchical and non-overlapping modular structures, the degeneracy problem is less severe and modularity maximization methods are likely to perform well; in other cases, modularity maximization can only provide a rough sketch of some parts of a network's modular organization". Thus, when one consider the large-scale networks with relatively complicated community structures, the degeneracy problem of modularity will become especially serious. This may be the reason why the performance of our MABA is not as good as some excellent algorithms (such as infomap and RN) in our experiment when analyzing large heterogeneous networks. For this problem, very recently, Khadivi et al. [9] proposed a solution which looks like reasonable. They adopted a weighting scheme prior to applying a community detection method based on modularity optimization. The weighting scheme should be able to discriminate between intercluster and intracluster edges. It should be able to strengthen the weights of intracluster edges while decreasing the weights of intercluster ones in order to mitigate the extreme degeneracy of modularity. Considering the discussion above, our future work can be laid as follows. We intend to improve our algorithm by designing a proper weighting scheme as its preprocessing step. The purpose is to make this new improved method well suitable for mitigating the degeneracy problem, and have stronger ability to deal with large heterogeneous networks.

Moreover, the configurations explored by our algorithm are those reachable by changing the groups node by node. This means that the initial conditions are an issue, and the exploration of the landscape of modularity $Q$ is partial. In our further work, we will improve the algorithm from these two aspects. First, we wish to present a new and effective initialization method, so as to make our ant-based algorithm not such sensitive as before to the initial condition. Then, we will also deliberately design a more powerful search strategy by adding some new manipulations rather than just moving nodes, purpose of which is to make our algorithm able to mitigate the partial exploration problem.

At last, it is obvious that our algorithm possesses the potential ability to uncover multi-scale hierarchical structures. Thus, in the experiment part, we also give a simple example to illustrate its capacity to unfold hierarchy. But we did not further offer this type of application on some large-scale real networks. It is known from Ref. [38] that complex networks in the real world often have natural and meaningful hierarchy. Therefore, in our future work, we also intend to further apply MABA to several real-world networks which have natural hierarchy, and wish to uncover and interpret the significative hierarchical community structures that is expected to be found in them.

**Acknowledgment**

This work was supported by National Natural Science Foundation of China (60873149, 60973088, 61133011) and Scholarship Award for Excellent Doctoral Student granted by Ministry of Education (450060454018).